# Photo-magnetization in two-dimensional sliding ferroelectrics

Jian Zhou[*]

*Center for Alloy Innovation and Design, State Key Laboratory for Mechanical Behavior of Materials, Xi'an Jiaotong University, Xi'an 710049, China*

[*]Email: jianzhou@xjtu.edu.cn

Abstract

Light-matter interaction is one of the key routes to understanding and manipulating geometric and electronic behaviors of materials, especially two-dimensional materials which are optically accessible owing to their high surface to volume ratio. In the current work we focus on the recently discovered two-dimensional sliding ferroelectric materials, in which the out-of-plane electric polarization can be switched with a small horizontal translation in one layer. Combining symmetry analysis and first-principles calculations, we predict that light illumination could inject non-equilibrium magnetic moments into the sliding ferroelectrics. Such magnetic moment is composed of both spin and orbital degrees of freedom contributions. We use $ZrI_2$, $WTe_2$, and $MoS_2$ bilayer ferroelectrics to illustrate our theory. Under intermediate light illumination, one can yield non-equilibrium magnetic moments on the order of $0.1 - 1$ $\mu_B$ in these systems, which also depends on the polarization nature of incident light. Furthermore, we show that such photo-injected magnetism changes its sign when the sliding dipole moment switches. This photo-magnetization can be detected by magneto-optical methods (such as Kerr or Faraday effect), which serves as an indicator of sliding ferroelectricity. Hence, one can use an all-optical pump and probe setup to measure and detect the subtle sliding ferroelectric phase.

**Introduction.**

Two-dimensional (2D) ferroelectric materials hold their potential applications in ultrafast information storage and data memory with low energy cost and high volumetric density[1-3]. During the past decade, most 2D sub-nanometer ferroelectric materials have been realized by physically exfoliating layered materials into their monolayer (or few layer thin film) form, which lacks inversion symmetry intrinsically and possesses an in-plane electric polarization or out-of-plane dipole moment. These materials, such as GeS[4-7], SnTe[8-10], $In_2Se_3$[11-14], $CuInP_2S_6$[15-17], 2D perovskite[18-20], and their analogous systems[21-23], are naturally van der Waals (vdW) layered materials without dangling bonds on their surface. This ferroelectricity nature have been demonstrated in both theoretical and experimental works.

In addition to these inherent ferroelectric 2D structures, very recently, Wu *et al.* have proposed another type of 2D ferroelectrics, based on the vertical stacking mismatch between two or more vdW layers[24]. The electric dipole moment is mainly along the out-of-plane (*z*) direction, and its flip can be realized by sliding the vdW layers of a short distance (usually on the order of $0.1 - 1$ Å). In this pioneering work, they showed that the AB stacking of honeycomb semiconducting sheets is a potential system, such as BN and $MoS_2$ bilayers. The vertically aligned B-N pair (or Mo-S dimer pair) produces an out-of-plane dipole moment, and a sliding of $a/\sqrt{3}$ (*a* being in-plane lattice constant) of one layer could flip the dipole. The energy cost for such transition is rather low (on the order of $0.1 - 1$ µJ/cm$^2$), compared with conventional bulk and other 2D ferroelectric materials. The different stacking patterns may serve as memory storage unit, which can be integrated and rolled up in practical usage[25]. Subsequently, another type of sliding ferroelectric (SFE) materials, $T_d$-$WTe_2$, has been theoretically predicted[26]. Even though each vdW layer of $WTe_2$ is centrosymmetric ($\mathcal{P}$), the stacking mismatch of the layer-resolved inversion center in the $T_d$ form yields a finite out-of-plane dipole moment. Theoretical calculations suggest that linearly polarized light (LPL) induced bulk photovoltaic effect, namely, shift current would flip its direction once the SFE transition occurs in the trilayer $WTe_2$, while the bilayer structure does not have such property[27]. Motivated by these studies, recently several groups have showed that $ZrI_2$ is another promising SFE material with semiconducting feature[28-30], which would be more suitable for nanoelectronic devices than the semi-metallic $WTe_2$.

The first experimental observation of 2D sliding ferroelectrics was reported in 2018 based on $WTe_2$[31]. This work is in parallel with theoretical suggestions. Then more and more experimental discoveries arise, invoked by the theoretical proposals. For example, Xiao *et al.* have demonstrated that Berry curvature changes its sign during the sliding of bulk $T_d$-$WTe_2$, similar as in trilayer $WTe_2$, so that the nonlinear Hall current (determined by Berry curvature dipole) could reverse under a small sliding of $WTe_2$ layer[32]. In 2021, two independent experiments have demonstrated that sliding of bilayer BN could yield such interfacial ferroelectricity[33,34]. Similar SFE in transition metal dichalcogenide (such as $MoS_2$) has also been realized in experiments recently[35].

Since the energy barrier separating the two SFE states (dipole moment up and down, denoted as $P_\Uparrow$ and $P_\Downarrow$, respectively) is low, one has to reduce direct and strong interactions between the tip and samples during its operation. Hence, non-contacting approaches would be more preferable for probing these states, compared with electrical or mechanical schemes which usually require direct contacts and may introduce unwanted impurities and disorders. Note that in the pioneer work[31], Fei *et al.* used graphene as a sensor to detect ferroelectric switch without contacts to the samples. Optical illumination, on the other hand, is a widely accepted method to detect and manipulate materials without direct contacts[7]. This concept has been considered in the $T_d$-$WTe_2$, in which photo-induced electric current direction reverses in the $P_\Uparrow$ and $P_\Downarrow$ states[27,32]. But measuring electric current (or voltage) also requires depositing electrodes at the sample edge. In addition, as the thinnest SFE system, bilayer $WTe_2$ does not possess such property – its photocurrent magnitude and direction remain unchanged under dipole flip. Hence, it would be ideal to resort to other all-optical method which can detect and distinguish bilayer SFE $P_\Uparrow$ and $P_\Downarrow$ states.

In this work, we predict that light irradiation induced static magnetism, namely,

photo-magnetization, could be a simple and addressable approach. By analyzing the symmetry of these two types of bilayer SFE materials, we show that light induced static magnetization can be opposite in the $P_\Uparrow$ and $P_\Downarrow$ states. The photo-magnetization is evaluated according to quadratic response theory[36]. Note that its linear response process describes current induced magnetization, namely, Edelstein effect[37-40]. Thus, this optical scheme is a second order nonlinear Edelstein (NLE) effect[41]. We use $ZrI_2$, $WTe_2$, and $MoS_2$ bilayers and perform first-principles density functional theory (DFT) to illustrate our theory. For the $ZrI_2$ and $WTe_2$, we suggest that one can apply circularly polarized light (CPL) and inject $y$-polarized magnetism. Flipping the SFE dipole moment would simultaneously reverse the $\langle \delta m^y \rangle$ magnetic moment. In detail, we also show that the photo-magnetizations are antiparallel in the two vdW layers, exhibiting ferrimagnetic alignment. Sliding of one layer would adjust their relative magnitude, thus reversing the net magnetization. As for the honeycomb $MoS_2$ bilayer, we demonstrate that LPL illumination could induce $\langle \delta m^x \rangle$ (along the zigzag direction) magnetization that is contrast in the $P_\Uparrow$ and $P_\Downarrow$ states. These photo-magnetization could be further probed and detected via other optical approaches, such as magneto-optical Kerr effect or Faraday effect. Thus, an all-optical pump (NLE process) and probe (Kerr or Faraday process) setup is appropriate to read the SFE states, which is non-contacting, non-invasive, and less susceptible to lattice damage in practical applications.

**Results.**

When light irradiates into a material, it excites electronic interband transitions between valence and conduction bands. In the quadratic optical process, the system absorbs one photon and emit another, leaving a perturbed state with magnetization change. This process can be expressed as

$$\langle \delta m^a \rangle(\omega = 0) = \text{Re} \sum_{\Omega = \pm \omega} \chi^a_{bc}(\Omega) E_b(\Omega) E_c^*(\Omega), \tag{1}$$

where $a, b, c$ are Cartesian coordinates and $\boldsymbol{E}(\omega)$ is the Fourier transformation of alternating electric field (light) with $\omega$ being angular frequency. The complex conjugate of electric field is $\boldsymbol{E}^*(\omega) = \boldsymbol{E}(-\omega)$. This equation explicitly indicates that under a two-photon (polarization along $b$ and $c$ respectively) process with opposite phase, the system could generate magnetization $\langle \delta m^a \rangle$ which points along direction-$a$. This photo-magnetization is static, i.e., its angular frequency is zero. According to previous works[38,41-46], this injected magnetization contains two parts, spin and orbital degrees of freedom. Hence, the response coefficient $\chi$ is a third-order tensor (with two subscripts indicating light polarization and the superscript referring to magnetization direction) and also contains spin and orbital contributions, $\chi(\omega) = \chi^S(\omega) + \chi^L(\omega)$. According to the second order Kubo response theory, the total NLE response function can be explicitly evaluated in the independent particle framework

$$\chi_{bc}^a(\omega) = -\frac{\mu_B e^2 S_{\text{u.c.}}}{\hbar^2 \omega^2} \int \frac{d^2\boldsymbol{k}}{(2\pi)^2} \sum_{O=S,L} g_O \sum_{mnl} \frac{f_{lm} v_{lm}^b}{\omega_{ml} - \omega + i/\tau} \left( \frac{O_{mn}^a v_{nl}^c}{\omega_{mn} + i/\tau} - \frac{v_{mn}^c O_{nl}^a}{\omega_{nl} + i/\tau} \right) \quad (2)$$

Here $S_{\text{u.c.}}$ is the area of a simulation unit cell, and integral is performed in the 2D first Brillouin zone. Operator $\hat{O}$ indicates spin ($S$) or orbital ($L$) angular momentum operator. It involves three band transitions among band-$l$, $m$, and $n$. We use velocity gauge with velocity matrix $\boldsymbol{v}_{lm} = \langle l | \hat{\boldsymbol{v}} | m \rangle$ and angular momentum matrix $O_{mn}^a = \langle l | \hat{O}^a | m \rangle$. $f_{lm} = f_l - f_m$ and $\omega_{ml} = \omega_m - \omega_l$ are occupation and frequency difference between bands $m$ and $l$. All the $\boldsymbol{k}$-dependence on the velocity, occupation, and frequency matrices are omitted. The Landé g-factor $g_O$ for spin and orbital degrees of freedom are 2 and 1, respectively. In order to phenomenologically incorporate environmental scatterings such as disorder, impurity, electron-phonon coupling, etc., we include carrier lifetime $\tau$. Note that this lifetime should be different for different states (band index $n$ and momentum $\boldsymbol{k}$), but a complete evaluation is impossible, even in a perfect crystal. Hence, we follow previous works[41,47,48] to adopt constant relaxation time approximation in the perturbation theory (taken to be 0.2 ps in this work). In addition, it has been experimentally shown that carrier lifetime can be longer than a few picoseconds[49], which suggests that 0.2 ps is a conservative value. Since we are treating 2D materials which is ultrathin in the out-of-plane dimension, local field effect would be marginal and independent particle approximation is appropriate. This NLE response function gives the photo-magnetization injected in one simulation unit cell. Considering the phase difference between the LPL and CPL, we can use the symmetric real and asymmetric imaginary parts of $\chi_{bc}^a(\omega)$ as

$$\eta_{bc}^a(\omega) = \frac{1}{2}\text{Re}(\chi_{bc}^a + \chi_{cb}^a), \tag{3a}$$

$$\xi_{bc}^a(\omega) = \frac{1}{2}\text{Im}(\chi_{bc}^a - \chi_{cb}^a). \tag{3b}$$

We will use them to evaluate the NLE responses under LPL and CPL, respectively. In order to distinguish the subscript between the LPL and CPL, we use ↺ and ↻ to denote the left-handed and right-handed CPL (instead of subscripts $xy$ and $yx$ in $\xi$), respectively. Since both spin and orbital are angular momenta, and they show same symmetry transformation under mirror $\mathcal{M}$, rotation $\mathcal{C}$, and time reversal $\mathcal{T}$. We will only discuss their total contributions to magnetization.

We now discuss NLE responses of ZrI$_2$ bilayers (Fig. 1). Each of these two layers shows the T′ phase, which is centrosymmetric. When they are vdW stacked, the whole system may take different patterning orders. The high symmetric structure belongs to 2D layer group of $Pm21b$, which corresponds to $C_{2v}$ point group at the origin [Fig. 1(b)]. In this structure, the two layers are related to each other by a glide operation, $\{\mathcal{M}_z|(0,\frac{1}{2})\}\mathbf{x}_{L1} = \mathbf{x}_{L2}$, where $\mathbf{x}_{L1}$ and $\mathbf{x}_{L2}$ are atomic coordinates of the lower (L1) and higher (L2) layers, respectively. This indicates that the centrosymmetric points of each layer are vertically aligned in $Pm21b$. Hence, this structure is still centrosymmetric without spontaneous polarization. However, according to previous works[28-30], this high symmetric $Pm21b$ is dynamically unstable and can lower its energy into two equivalent structures, which belong to 2D layer group of $Pm11$ (point group of $C_s$). The sliding distance of the L2 is 0.28 Å from the saddle (SD) $Pm21b$, and the energy gain is 1.50 meV per formula unit (f.u.), or 0.4 μJ/cm². One thus obtains $P_⇑$ and $P_⇓$ states. In the $P_⇓$ structure [Fig. 1(a)], the two vdW layers can be transformed via $\{\mathcal{M}_z|(0,t_b)\}\mathbf{x}_{L1} = \mathbf{x}_{L2}$ with $t_b = 0.455$. Similarly, for the $P_⇑$ [Fig. 1(c)], it becomes $\{\mathcal{M}_z|(0,1-t_b)\}\mathbf{x}_{L1} = \mathbf{x}_{L2}$. Even though both $P_⇓$ and $P_⇑$ are achieved by a translation of the L2 layer, one can also connect them by a 180° rotation, $\mathcal{C}_{2y}\mathbf{x}_{P_\Downarrow} = \mathbf{x}_{P_\Uparrow}$. Note that all of these three states contain a $\mathcal{M}_x$ mirror symmetry.

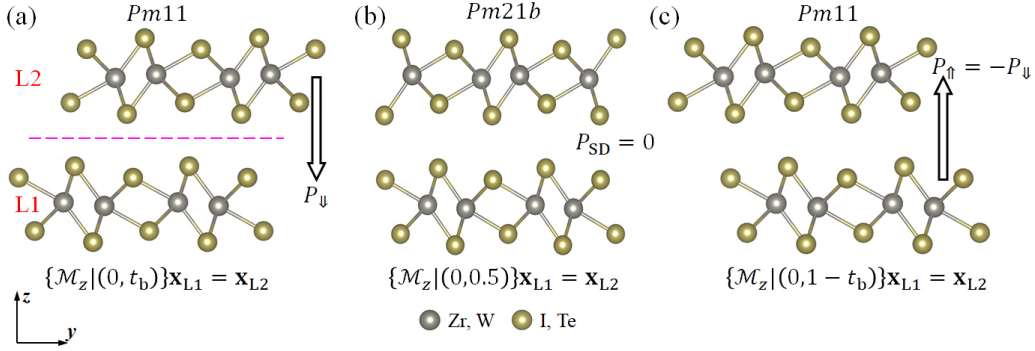

**Fig. 1 Atomic geometry and layer group of bilayer ZrI$_2$ (and WTe$_2$).** (a) $P_\Downarrow$, (b) saddle point with $P_{SD} = 0$, and (c) $P_\Uparrow$ states. The atomic coordinates in the two layers (L1 and L2) can be transformed by a $\mathcal{M}_z$ mirror [dashed line in panel (a)] with a translation ($t_b = 0.455$), as denoted below the structures.

These spatial symmetry imposes constraints on the NLE response functions. We denote the numerator of the integrand in Eq. (2) as $N^{a,bc}(\mathbf{k}) = O^a v^b v^c(\mathbf{k})$. Since the $\mathcal{M}_x$ is always present in all three states, we first analyze its effect. After $\mathcal{M}_x$ reflection, the $\mathbf{k} = (k_x, k_y)$ becomes $\mathcal{M}_x \mathbf{k} = \widetilde{\mathbf{k}} = (-k_x, k_y)$. The velocity operator transforms as $\mathcal{M}_x v^i = (-1)^{\delta_{xi}} v^i$ (where $\delta_{xi}$ is Kronecker delta, $i = x, y$). The angular momentum $O$ transforms as pseudovector, which satisfies $\mathcal{M}_x O^i = -(-1)^{\delta_{xi}} O^i$. We thus have $\mathcal{M}_x N^{a,bc}(k_x, k_y) = -(-1)^{\delta_{xb}+\delta_{xc}+\delta_{xa}} N^{a,bc}(-k_x, k_y)$. The denominator in Eq. (2) is only composed of frequency, which is even under $\mathcal{M}_x$. Therefore, under in-plane polarized CPL irradiation ($b = x$ and $c = y$), magnetization along-$x$ will be symmetrically forbidden as its integration over the Brillouin zone is always zero. Our numerical results also confirm this symmetry arguments.

Next, we focus on the $\langle \delta m^y \rangle$ that is nonzero under CPL irradiation. In the time-reversal symmetric $\mathcal{T}$ systems, previous works[41,45,50] have shown that the response function in Eq. (2) can be reduced into a two-band expression (under the assumption of long carrier lifetime limit, $\tau \to \infty$)

$$\xi^a_{bc}(\omega) = -\tau \frac{\pi \mu_B e^2 S_{\text{u.c.}}}{2\hbar^2} \int \frac{d^2 \mathbf{k}}{(2\pi)^2} \sum_{O=L,S} g_O \sum_{ml} f_{lm} [r^b_{lm}, r^c_{ml}] (O^a_{mm} - O^a_{ll}) \delta(\omega_{ml} - \omega).$$

(4)

Here the commutator $[r_{lm}^b, r_{ml}^c] = r_{lm}^b r_{ml}^c - r_{lm}^c r_{ml}^b = -i\epsilon_{bcd}\Omega_{lm}^d$ corresponds to interband Berry curvature $\Omega_{lm}^d$, and $(O_{mm}^a - O_{ll}^a)$ measures the angular momentum difference between band $m$ and $l$. The Eq. (4) describes a physical process that CPL excites electron from valence to conduction band according to transition rate $R \propto [r_{lm}^b, r_{ml}^c]\delta(\omega_{ml} - \omega)$, and the angular momentum difference gives its moment injection $\langle\delta m\rangle \propto \xi$. The pumping is compensated by the relaxation time $\tau$. This is a saturation time, beyond which no further magnetism injection exists. In order to evaluate layer contributions, we introduce a projector operator $\hat{Q}_j = \sum_{i \in \{L_j\}} |\psi_i\rangle\langle\psi_i|$ where $|\psi_i\rangle$ represent Wannier functions centered at the $j$-th layer ($j = 1, 2$). Then we multiply the angular momentum operators in Eqs. (2) and (4) by the projector, namely, $\hat{O} \to \hat{O}\hat{Q}_j$. This would yield angular momentum localized on the layer-$j$.

We plot the band structure of $P_\Uparrow$-ZrI$_2$ bilayer in Fig. 2(a), with the intrinsic $\langle m^y\rangle$ of each state represented by different colors. One sees that consistent with previous calculations[30], the ZrI$_2$ is a semiconductor with bandgap of ~0.3 eV. Therefore, when the incident light energy $\hbar\omega$ is below 0.3 eV, no magnetism injection occurs [Fig. 2(b)−(d)]. We only show our results for $\xi_\circlearrowleft^y(\omega)$, which represents photo-magnetization under left-handed CPL. The right-handed CPL produces opposite magnetisms as $\xi_\circlearrowright^y(\omega) = -\xi_\circlearrowleft^y(\omega)$. This can be clearly seen from the position commutation in Eq. (4). From Fig. 2(b), one observes that $\xi_\circlearrowleft^y(\omega)$ for $P_\Uparrow$-ZrI$_2$ is on the order of 0.1 $\mu_B \times$ nm$^2$/V$^2$. This indicates that a CPL laser with $10^{11}$ W/cm$^2$ intensity (~1 V/nm) could generate magnetism on the order of 0.1 $\mu_B$. The largest NLE response occurs at incident energy of $\hbar\omega = 0.6$ eV with total $\xi_{xy}^y = 0.58$ $\mu_B \times$ nm$^2$/V$^2$. Compared with previous works on static electric field driven magnetization (linear Edelstein effect, $\langle\delta m\rangle \propto E^1$)[38,40], this is magnetization injection is $1 - 2$ orders of magnitude larger under the same electric field magnitude. We find that in most incident energy regime, the two layers contribute opposite magnetizations, along $y$ and $-y$. For example, when $\hbar\omega = 0.75$ eV, the NLE response function is 0.4 and −0.1 $\mu_B \times nm^2/V^2$ in the L1 and L2, respectively. This shows a ferrimagnetic alignment. We plot the $\bm{k}$-distribution of the integrand of Eq. (4) in Figs. 2(e) and (f) for each layer, which illustrates the momentum space contribution. One can observe that the distribution obeys $\mathcal{M}_x$ reflection, consistent with our previously symmetry analysis. The main peaks lie near the $k_x = \pm 0.3$ Å$^{-1}$ lines, while other regions contribute marginally as their bandgap is generally larger than incident photon energy.

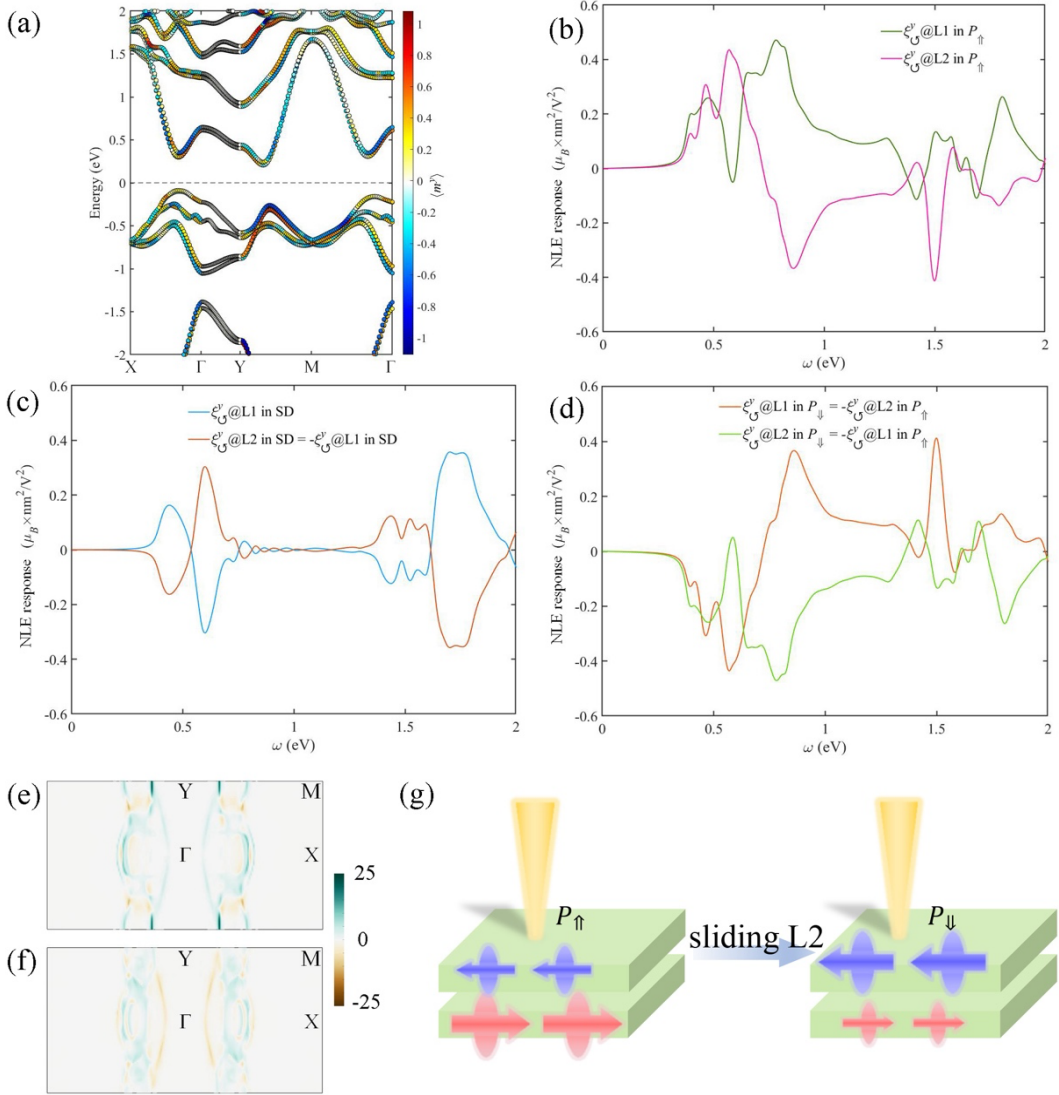

**Fig. 2 Photon induced magnetization along $\bm{y}$ of bilayer ZrI$_2$.** (a) Band structure of $P_\Uparrow$-ZrI$_2$ with colors representing $\langle m^y \rangle = 2\langle S^y \rangle + \langle L^y \rangle$ of each state (in $\mu_B$). (b)−(d) Layer-resolved nonlinear Edelstein response function $\xi^y_\circlearrowleft(\omega)$ for $P_\Uparrow$, SD, and $P_\Downarrow$-ZrI$_2$, respectively. (e) and (f) Brillouin zone distribution of $\xi^y_\circlearrowleft(\omega = 0.75$ eV$)$ for $P_\Uparrow$-ZrI$_2$ in the L1 and L2, respectively. (g) Schematic plot of photo-magnetization in bilayer ZrI$_2$ during layer sliding. The contrasting magnetic moment states are denoted by arrows with different sizes and colors, which can be probed by magneto-optical methods. Note that in this theory, vertical incidence of light is assumed so that the electric field components lie in the $xy$ plane, rather than oblique incidence.

When the L2 layer slides 0.28 Å, the system reaches high symmetric SD structure. Figure 2(c) plots the NLE induced $\langle \delta m^y \rangle$ in each layer. It shows exact antiferromagnetic pattern, i.e., they align oppositely with the same magnitude under any specific incident energy. Hence, the net photo-magnetization is zero. This is because the system now contains a glide mirror symmetry, $\{\mathcal{M}_z | (0, \frac{1}{2})\}$. Hence, we show that a small distance translation could drastically change the NLE response, so that this photo-magnetization could serve as a sensitive method to detect vdW sliding, which only demands marginal energy with very small geometric variation.

Furthermore, another 0.28 Å sliding would bring the system into $P_\Downarrow$-ZrI$_2$ structure. Our results show that the NLE response $\xi_\circlearrowleft^y(\omega)$ is opposite to that in $P_\Uparrow$-ZrI$_2$. In detail, the $\xi_\circlearrowleft^y(\omega)$@L1 in $P_\Uparrow$ equals to $-\xi_\circlearrowleft^y(\omega)$@L2 in $P_\Downarrow$, and $\xi_\circlearrowleft^y(\omega)$@L2 in $P_\Uparrow$ equals to $-\xi_\circlearrowleft^y(\omega)$@L1 in $P_\Downarrow$. Hence, the total photo-magnetization $\langle \delta m^y \rangle$ also flips its sign from $P_\Uparrow$ to $P_\Downarrow$. This can be understood by noting that $\mathcal{C}_{2y} \mathbf{x}_{P_\Downarrow} = \mathbf{x}_{P_\Uparrow}$. Again we look at the numerator in Eq. (2). It can be easily shown that $\mathcal{C}_{2y} N^{y,xy}(\mathbf{k}) = \mathcal{M}_x \mathcal{M}_z N^{y,xy}(\mathbf{k}) = -N^{y,xy}(\widetilde{\mathbf{k}})$. The denominator is only composed of eigenenergies which follow $\mathcal{C}_{2y}$. Therefore, the $\mathbf{k}$-space integration gives that photo-magnetization $\xi_\circlearrowleft^y(\omega)$ in $P_\Uparrow$ is opposite of that in $P_\Downarrow$. In particular, one notes that the operation $\mathcal{C}_{2y}$ maps the L1 in $P_\Uparrow$ to L2 in $P_\Downarrow$ (L2 in $P_\Uparrow$ to L1 in $P_\Downarrow$), so the layer contribution also flips. These are consistent with our numerical calculations. We schematically plot this sliding induced layer-magnetization in Fig. 2(g). The arrow size and color represent the magnetization and direction, respectively. Under intermediate pumping CPL intensity (~$10^{10}$ W/cm$^2$), the photo-magnetization in the $P_\Uparrow$ and $P_\Downarrow$ states can differ by 0.06 $\mu_B$, which is observable experimentally.

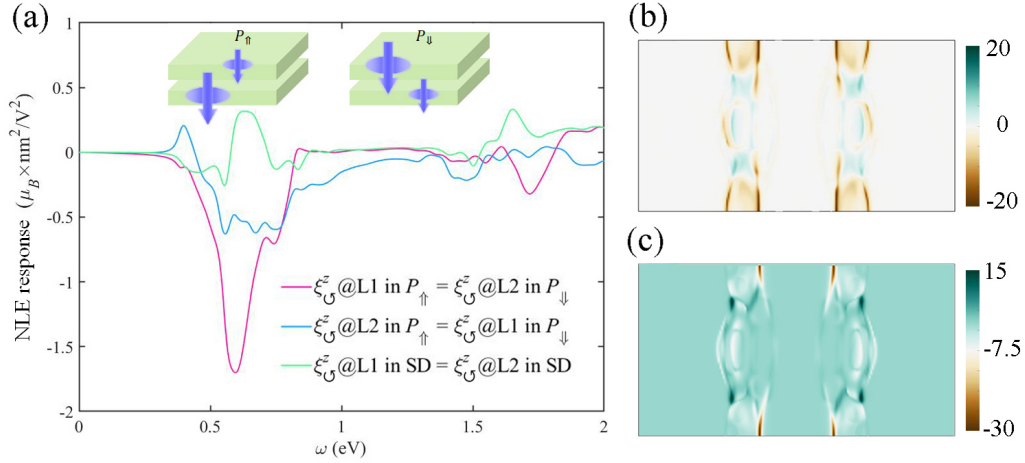

**Fig. 3 Photo-magnetization induced magnetic moment along $z$ in bilayer ZrI$_2$.** (a) Layer-resolved $\xi_G^z(\omega)$ for the $P_\Uparrow$, SD, and $P_\Downarrow$, which shows a general staggered magnetic moment distribution in the two layers (inset). (b) and (c) show the Brillouin zone distribution of $\xi_G^z(\omega = 0.7\ eV)$ for L1 and L2 in $P_\Uparrow$, respectively.

We also evaluate the injected magnetization along the $z$-direction. The results are shown in Fig. 3. We find that the total NLE responses in $P_\Uparrow$ to $P_\Downarrow$ are the same and could reach $-2.4\ \mu_B \times nm^2/V^2$, while the SD structure gives smaller NLE response function. Specifically, when the incident photon energy is in the $0.4 - 0.8$ eV, the L1 hosts larger (more negative) photo-injected magnetic moment $\langle \delta m^z \rangle$ than the L2. We plot the $\boldsymbol{k}$-resolved contribution at an incident energy of $\hbar\omega = 0.7$ eV, in Figs. 3(b) and (c). Similarly as in previous case, the main contribution is around $k_x = \pm 0.3\ Å^{-1}$ and the $\mathcal{M}_x$ symmetry is still preserved. Sliding from $P_\Uparrow$ to SD would make the photo-magnetizations in the two layers equal, and further sliding flips their relative magnitudes. Unlike the $\langle \delta m^y \rangle$ case (Fig. 2), here the $\xi_G^z(\omega)$@L1 in $P_\Uparrow$ equals to $\xi_G^z(\omega)$@L2 in $P_\Downarrow$, and $\xi_G^z(\omega)$@L2 in $P_\Uparrow$ equals to $\xi_G^z(\omega)$@L1 in $P_\Downarrow$. This is consistent with the even symmetry of $N^{z,xy}(\boldsymbol{k})$ under $\mathcal{C}_{2y}$. Therefore, the $z$-component of NLE magnetism cannot distinguish the two states in SFE ZrI$_2$ bilayer.

Then we briefly discuss WTe$_2$ bilayer. Note that WTe$_2$ is semimetallic without a finite bandgap, here we also include intraband contribution from the Fermi surface

$$\chi_{bc}^{a,\text{FS}}(\omega) = -\frac{\mu_B e^2 S_{\text{u.c.}}}{\hbar \omega^2}$$

$$\times \int \frac{d^2\mathbf{k}}{(2\pi)^2} \sum_{O=S,L} g_O \left[ \sum_{mn} O_{nn}^a \frac{v_{nm}^b v_{mn}^c}{\omega_{mn}-\omega} + \sum_n O_{nn}^a w_{nn}^{bc} \right]. \quad (5)$$

Here $w_{nn}^{bc} = \langle n | \frac{\partial^2 H}{\hbar \partial k_b \partial k_c} | n \rangle$ and $E_F$ is the Fermi energy. The Dirac delta function $\delta(\hbar\omega_n - E_F)$ indicates Fermi surface contribution. This Fermi surface effect mainly comes from the off-diagonal element contribution in the perturbed density matrix, and it has been applied to evaluate photocurrent in metals[51]. Our calculations show that compared with interband NLE response [as in Eqs. (2) and (4)], the intraband contribution is much smaller (2 – 3 orders of magnitude smaller). Figure 4 shows the total (interband and intraband) NLE response $\xi_O^y(\omega)$ of each layer for the $P_\Uparrow$ and $P_\Downarrow$-WTe$_2$ bilayers. Since the symmetry arguments for ZrI$_2$ bilayer are the same for WTe$_2$, their layer dependent $\xi_O^y(\omega)$ relations also agree. However, one sees that the magnitude of NLE responses are much larger for WTe$_2$, which reaches ~4 $\mu_B \times$nm$^2$/V$^2$ at an incident photon energy of 0.1 eV. In detail, for $P_\Uparrow$-WTe$_2$, the magnetism in L1 points toward $-y$ ($\langle \delta m^y \rangle = -4.5$ $\mu_B \times$ nm$^2$/V$^2$), while the L2 holds $\langle \delta m^y \rangle = 8.5$ $\mu_B \times$ nm$^2$/V$^2$, along $+y$. Therefore, a laser with its intensity of 3.2×10$^9$ W/cm$^2$ is enough to inject 0.1 $\mu_B$ magnetic moment, which is sufficiently large for detection. Flipping the SFE through a ~0.7 Å sliding of L1 could yield a 0.2 $\mu_B$ moment change. We plot their NLE distributions in the $\mathbf{k}$-space, which show dominant peaks at $k_x = \pm 0.2$ Å. Again, one has to note that the above calculations and discussions are based on left-handed CPL irradiation. If right-handed CPL is used, all signs will be reversed, which will not be discussed in detail.

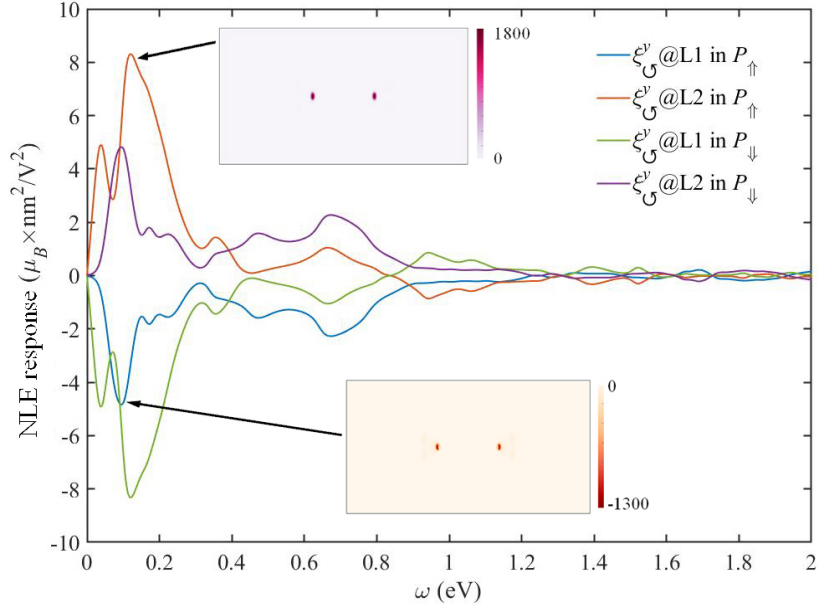

**Fig. 4 Layer-resolved NLE response function $\xi_\circlearrowleft^y(\omega)$ of bilayer WTe$_2$ for $P_\Uparrow$ and $P_\Downarrow$ states.** Both interband and intraband contributions are included. Insets show the *k*-distributions for the two layers at $\omega = 0.1$ eV for $P_\Uparrow$-WTe$_2$.

We also consider another type of SFE material, bilayers of honeycomb lattice. We use transition metal dichalcogenide MoS$_2$ as the exemplary system. The high symmetry MoS$_2$ bilayer stacking is an eclipsed pattern where L2 is fully on top of L1, usually denoted as AA stacking (layer group of $P\bar{6}m2$). However, this stacking pattern is dynamically unstable, and it would spontaneously slide into two structures with nonzero out-of-plane dipole moment. Their atomic geometries are shown in Fig. 5(a), and they belong to layer group of $P3m1$. All these systems possess mirror $\mathcal{M}_x$ symmetry ($x$ denotes the zigzag direction). In the $P_\Uparrow$-MoS$_2$, the atomic coordinates in two layers are related by a glide operation, $\{\mathcal{M}_z | \left(\frac{2}{3}, \frac{1}{3}\right)\} \mathbf{x}_{L1} = \mathbf{x}_{L2}$ or $\{\mathcal{M}_z | \left(\frac{1}{3}, \frac{2}{3}\right)\} \mathbf{x}_{L2} = \mathbf{x}_{L1}$. Similar argument holds for the $P_\Downarrow$-MoS$_2$. We can also use mirror symmetry to connect these two SFE states, namely, $\mathcal{M}_z \mathbf{x}_{P_\Uparrow} = \mathbf{x}_{P_\Downarrow}$. This is illustrated in Fig. 5(a). The band structure of $P_\Uparrow$-MoS$_2$ is plotted in Fig. 5(b), which shows a large bandgap. One can also see clear valley splitting (under spin-orbit coupling) in the valence and conduction bands (at K and −K), and all the four valence valley states are singly degenerate. This is due to the dipole moment in $P_\Uparrow$ that create a build-in potential gradient, so that the L1 contributed states are energetically higher.

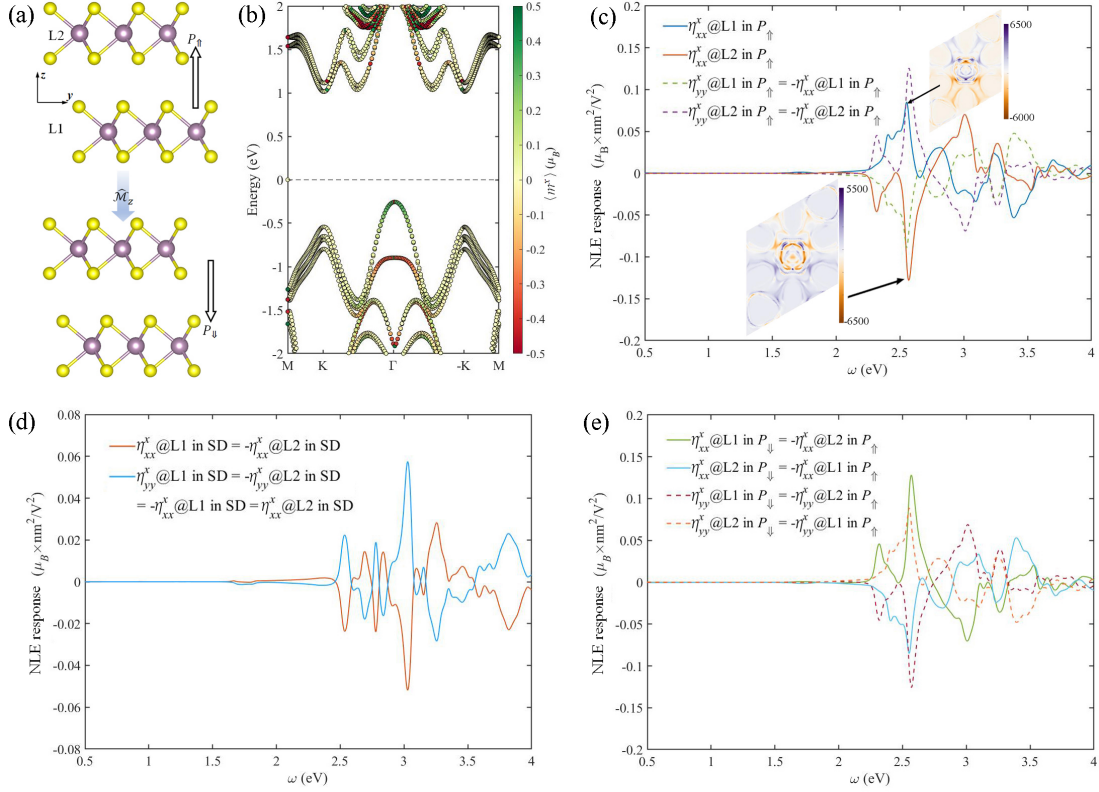

**Fig. 5 Photo-magnetization of bilayer sliding ferroelectric MoS$_2$.** (a) Atomic structures of bilayer MoS$_2$ in the $P_\Uparrow$ and $P_\Downarrow$ states. (b) Electron band dispersion along the high symmetric $\boldsymbol{k}$-path for $P_\Uparrow$-MoS$_2$, with colormap representing intrinsic magnetization of each state along $x$, $\langle m^x \rangle = 2\langle S^x \rangle + \langle L^x \rangle$. (c)−(e) Layer-resolved NLE response functions $\eta_{ii}^x(\omega)$ ($i = x, y$) under LPL for the $P_\Uparrow$, SD, and $P_\Downarrow$ configurations, respectively. $\boldsymbol{k}$-resolved NLE responses (at incident energy of $\hbar\omega = 2.6$ eV) for $P_\Uparrow$-MoS$_2$ under $x$-LPL is plotted as inset of (c).

Owing to the high symmetry of MoS$_2$ bilayer, the CPL injected photo-magnetization cannot provide opposite signatures for the $P_\Uparrow$ and $P_\Downarrow$ states. We thus consider the LPL injected magnetic moment for MoS$_2$ bilayer, according to Eq. (2). One may wonder that unlike CPL, the LPL does not break time-reversal symmetry $\mathcal{T}$, hence it seems counterintuitive that LPL could inject magnetization. This can be understood that energy dissipation from damping effect (finite carrier lifetime $\tau$) breaks $\mathcal{T}$. Here only photo-magnetization along $x$ ($\langle \delta m^x \rangle$) is nonzero, while the other two directions are symmetrically forbidden. In Figs. 5(c)−(e) we plot NLE response functions for the $P_\Uparrow$, SD, and $P_\Downarrow$ structures, respectively. The presence of $\mathcal{C}_{3v}$ symmetry yields $\eta_{xx}^x(\omega) = -\eta_{yy}^x(\omega)$. Hence, we will only focus on the *x*-LPL irradiation, and the *y*-LPL irradiation results would be similar (with a sign flip). In the $P_\Uparrow$-MoS$_2$ state, one clearly observes antiparallel pattern of NLE magnetization in the two layers when the incident photon energy is between 2.3 − 2.6 eV. Their magnitudes are not the same due to the presence of vertical dipole moment which breaks $\mathcal{M}_z$. The peak values for them are 0.08 (at 2.55 eV) and −0.13 (at 2.57 eV) $\mu_B \times$nm$^2$/V$^2$, respectively. Their Brillouin zone distributions are plotted as insets of Fig. 5(c). The $\mathcal{M}_x$ symmetry is kept, and the main contributions are around the Γ point. At the SD structure (which is energetically higher by 31 meV/f.u., or 11.4 μJ/cm$^2$), the NLE magnetization in the two layers are exactly antiferromagnetically aligned. Finally, the $P_\Downarrow$-MoS$_2$ is oppositely to the $P_\Uparrow$-MoS$_2$ state, with all magnetizations are flipped. Therefore, once again, the two polarization states of SFE MoS$_2$ bilayer show opposite NLE responses, which can be used to distinguish the SFE states. These two SFE states are separated by 1.82 Å sliding of one layer.

In order to see the ferroelectric switch effect clearly, we add the layer-resolved responses together (L1 and L2) and plot their NLE responses in Fig. 6. The results are well consistent with direct evaluations from Eq. (2) and Eq. (4) without layer projection scheme. These can be compared in experimental observations. One can see that the SD configuration always yields zero net NLE responses, owing to their high symmetry nature.

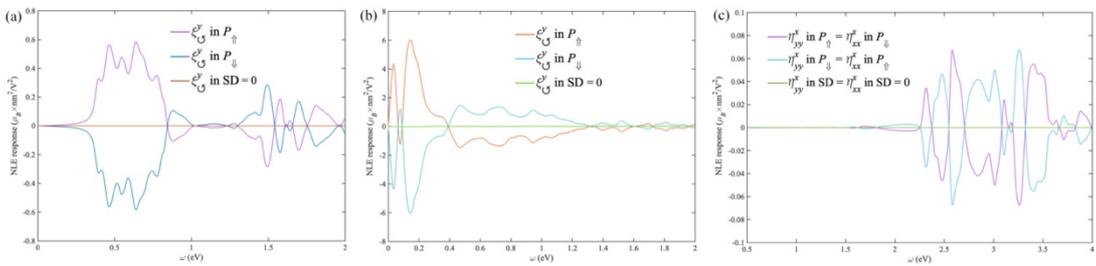

**Fig. 6 Total NLE response functions.** The layer summation of calculated NLE responses for bilayers (a) ZrI$_2$, (b) WTe$_2$, and (c) MoS$_2$.

**Discussion.**

Before concluding, we would like to briefly discuss the microscopic picture of NLE process and other SFE materials. One has to note that this photo-magnetization is actually non-equilibrium angular momentum (spin and orbital) injection, which is different from spin polarization created by a static magnetic field in equilibrium. This spin (or orbital) injection field essentially corresponds to a time-varying magnetic field, $\frac{\partial B}{\partial t}$[52]. For the CPL irradiation (onto time-reversal symmetric systems), Eq. (4) provides a clear microscopic picture of photo-magnetization process. The spin (orbital) injection originates from the spin (orbital) polarization difference at the valence and conduction band. This is akin to the injection current process[45,53], and it saturates at the lifetime $\tau$. Hence, it strongly depends on the sample quality. On the other hand, there is no clear relationship between $\tau$ and the LPL injected magnetization. One can look at the numerator and the denominator for NLE response function separately. The numerator (under LPL with polarization along $b$) is $N^{i,bb} = v^b v^b O^i$, and the first denominator is $D_1 = \frac{1}{\omega_{ml}-\omega+i/\tau} \frac{1}{\omega_{mn}+i/\tau}$ (the other denominator would be similar). Since the system is intrinsically $\mathcal{T}$-symmetric, one has $\mathcal{T} N^{i,bb}(\boldsymbol{k}) = -N^{i,bb,*}(-\boldsymbol{k})$ where $^*$ indicates complex conjugate. Hence, under integration over the whole Brillouin zone, only the imaginary part of the numerator $i \times \mathrm{Im} N$ contributes to the response function. Since the incident LPL does not contain a complex phase, it is the imaginary part of the denominators $D_1$ (and $D_2$) that multiplies with $i \times \mathrm{Im} N$ and gives the real response function. According to Sokhotski–Plemelj formula (when lifetime is sufficiently long, $\tau \to \infty$ ), we have $\mathrm{Im} D_1 = -\mathcal{P}\frac{\pi}{\omega_{ml}-\omega}\delta(\omega_{mn}) - \mathcal{P}\frac{\pi}{\omega_{mn}}\delta(\omega_{ml}-\omega)$ , where $\mathcal{P}$ denotes Cauchy principal value. In this case, the $\chi$ is independent on $\tau$. This is similar as the shift current generation, namely, LPL irradiated bulk photovoltaic effect for $\mathcal{T}$-symmetric systems[54]. The shift current that is usually described by a length-gauge two-band model is valid when we have long lifetime $\tau \to \infty$[55]. Under such assumption the shift current depends on $\tau^0$. On the other hand, for short lifetime, the dependence becomes complicated and $\chi$ generally reduces its overall magnitude. Our test calculations show that the NLE response functions are almost unchanged when $\tau$ is longer than 0.2 ps.

Our current work focuses on a few SFE bilayers and proposes their dipole moment dependent NLE responses. One has to be careful for its generalization into trilayer system, or even their bulk form. The response difference for the $P_⇑$ and $P_⇓$ states is determined by their own symmetry and operation to convert one to the other. In trilayer (odd-layered) and bulk forms, such operation may be different from the bilayer (even-layered) cases. Hence, the light polarization state and the injected magnetic moment direction can be different, even for the materials with same chemical formula.

Another interesting question would be if the NLE response flip can be seen in conventional ferroelectrics. We note that the linear (inverse) Edelstein effect, which is one of the key approaches for spin-charge conversion, shows electric polarization dependence in ferroelectrics[56,57]. However, for the light injected magnetization process (such as NLE in the present work), the symmetry requirement is different from linear Edelstein effect (which is under static electric field). In a lot of conventional ferroelectric materials, such as $BaTiO_3$ perovskite, they usually belong to a higher space group symmetry than 2D SFEs. The $P_⇑$ and $P_⇓$ states are usually connected via inversion symmetry $\mathcal{P}$ ($\mathcal{P}\mathbf{x}_{P_⇑} = \mathbf{x}_{P_⇓}$). If we examine the response function $\chi$, it is even under $\mathcal{P}$. Hence, the polarization switch does not flip photo-magnetization in these ferroelectrics. Even though this is not always the case in conventional ferroelectrics, polarization dependent photo-magnetization only exists when the materials and experimental setups are carefully designed. In the 2D sliding ferroelectric materials, the reduction in the out-of-plane dimension downgrades the space group symmetry into layer group symmetry, yielding more symmetry allowed photo-magnetization situations. For example, in the current study, the two vdW layers are connected by either $C_2$ rotation or mirror reflection $\mathcal{M}$, which allows photo-magnetization flipping under dipole switch. Hence, we suggest that the NLE effect is an appropriate way to be observed in SFE over conventional ferroelectrics.

One may wonder if optical irradiation could also switch the sliding polarization. In order to discuss this, let us briefly consider symmetry analysis. The hitherto discussed sliding ferroelectrics are mainly out-of-plane (z) polarized, and the flip between $P_⇑$ and $P_⇓$ is subject to a $\mathcal{C}_2$ rotation or mirror reflection $\mathcal{M}$. However, in most cases, the light alternating electric field does not break such symmetry. Therefore, under light irradiation, the Gibbs free energies of both $P_⇑$ and $P_⇓$ remain degenerate[10] and the dipole switch does not easily occur. Nevertheless, one can consider two possible approaches. One is to design 2D sliding ferroelectric materials that is out-of-plane symmetry broken, such as using 2D Janus TMD systems[58]. Hence, the $P_⇑$ and $P_⇓$ would become different in their configurations, and their optical responses are different, yielding different free energy. The other potential approach is to break time-reversal symmetry, which could be achieved via damping effect during transition. For example, it has been proved that optical excitation could trigger ferroelectric phase transition in $BaTiO_3$, even though the excited energy of the two ferroelectric phases are degenerate[59].

We expect this photo-magnetization in SFE can be observed in experiments due to the following reasons: (1) The NLE is a local effect which is evaluated in each unit cell. Hence its characteristic length scale can be as short as a few Å. This is different from nonlocal responses such as electric current generation, which depends on ferroelectric domain distributions. Since the energy barrier of coherent SFE switch is low, it can be anticipated that domain wall formation energy is also small, which corresponds to stacking fault and may occur with rippling of one layer. Therefore, the photo-magnetization may be advantageous than nonlocal effects. (2) According to our estimate, the photo-magnetization can be on the order of $0.1 - 1$ $\mu_B$ per unit cell. This magnitude is large enough to be observed in experiments. (3) In order to obtain observable magnetization, intermediate laser pump with its intensity on the order of $10^{10}$ W/cm$^2$ is required. Such laser intensity is not too strong to damage the sample, especially when the sample is in 2D nature, which significantly reduces direct laser absorption.

In conclusion, we use nonlinear optics theory to predict light injecting magnetization in SFE materials, which can be referred to as NLE responses. We use $ZrI_2$, $WTe_2$ and $MoS_2$ bilayers to show that CPL or LPL irradiation could trigger observable magnetic moments in their vdW layers. The stacking pattern could serve as an effective tuning parameter to modulate the NLE response functions. Particularly, we apply symmetry analysis and first-principles density functional theory calculations to suggest that this photo-magnetization can be opposite in the two SFE states. Even though the $P_\Uparrow$ and $P_\Downarrow$ states are spatially close (~1 Å sliding of one layer) and energetically degenerate, this photo-magnetization can be used as a sensitive method to probe and distinguish them. Unlike measuring nonlinear Hall conductance,[32] this optical approach is immune to direct contacts of tips or electrodes, so that the environmental influence can be greatly minimized.

**Methods.** Our first-principles calculations are based on density functional theory with its exchange correlation interaction treated in the form of Perdew-Burke-Ernzerhof (PBE) form[60], as implemented in the Vienna *ab initio* simulation package (VASP)[61]. Projector augmented wave method[62] is used to treat the core electrons, and the valence electrons are represented by planewave basis set with a kinetic cutoff energy set to be 400 eV. This approach would yield similar computational quality compared with all-electron calculations. Periodic boundary condition is used, which includes a vacuum space of ~15 Å in the out-of-plane $z$-direction, to eliminate the artificial interaction between the periodic images. Dipole corrections are adopted[63], to reduce the long range electrostatic interactions. Spin-orbit coupling (SOC) effect is incorporated in all calculations self-consistently. We use Monkhorst-Pack ***k*** mesh scheme[64] with grid density higher than $2\pi \times 0.02$ Å$^{-1}$ in the $x$ and $y$ plane. Convergence criteria of total energy and force component are set as $1\times 10^{-7}$ eV and $1\times 10^{-3}$ eV/Å, respectively. In order to fit the Hamiltonian, we use Wannier function representation, as implemented in the Wannier90 code[65]. The orbital angular momentum is evaluated using atomic orbitals, while inter-site contributions are omitted. In order to perform $k$-space integration for the nonlinear Edelstein response function, we use (501×301) and (601×601) grid meshes for $ZrI_2$ ($WTe_2$) and $MoS_2$ bilayers. The convergence of such computational procedure has been carefully tested.

**Acknowledgments.** This work was supported by the National Natural Science Foundation of China (NSFC) under Grant Nos. 21903063 and 11974270. J.Z. thanks valuable discussions with M. Wu. The calculations are performed in the HPC platform of Xi'an Jiaotong University.

**Data Availability.** All data generated or analyzed during this study are included in this published article, and are available from the authors upon reasonable request.

**Code Availability.** The related codes are available from the corresponding author upon reasonable request.

**Author Contributions.** J.Z. conceived the concept, performed calculations, analyzed the data, and wrote the manuscript.

**Competing Interests.** The author(s) declare no competing interests.